\documentclass[aps,prd,amssymb,cite,
amsfonts,epsf,preprintnumbers,nofootinbib,superscriptaddress]{revtex4}

\usepackage{kotex}
\usepackage[dvips]{graphicx}
\usepackage{bm,latexsym,amsmath,amssymb}
\usepackage[usenames,dvipsnames]{color}
\usepackage[colorlinks=true,linkcolor=blue]{hyperref}
\usepackage{soul}
\usepackage{epsfig}
\usepackage[mathscr]{eucal}
\usepackage{cancel}
\usepackage{mathrsfs}
\usepackage{pgf,tikz}
\usetikzlibrary{arrows,automata}
\usepackage{tcolorbox}

\definecolor{mypink1}{rgb}{0.858, 0.188, 0.478}
\definecolor{mypink2}{RGB}{219, 48, 122}
\definecolor{mypink3}{cmyk}{0, 0.7808, 0.4429, 0.1412}
\definecolor{mygray}{gray}{0.6}
\definecolor{pptbg}{rgb}{0.961,0.945,0.863}

\newcommand{\be}[1]{\begin{equation} \label{#1}}
\newcommand{\ee}{\end{equation}}
\newcommand{\bex}{\begin{equation*}}
\newcommand{\eex}{\end{equation*}}
\newcommand{\bea}{\begin{eqnarray}}
\newcommand{\eea}{\end{eqnarray}}
\newcommand{\ba}{\begin{array}}
\newcommand{\ea}{\end{array}}
\newcommand{\nn}{\nonumber}

\newcommand{\bel}{\begin{align}}
\newcommand{\eel}{\end{align}}

\newcommand{\vi}{U_{\rm i}}

\begin{document}
\title{Motions of a billiard ball after a cue stroke}

\author{Hyeong-Chan Kim}
\affiliation{School of Liberal Arts and Sciences, Korea National University of Transportation, Chungju, 27469, Korea}
\email{hckim@ut.ac.kr}

\begin{abstract}
We study the collision between the cue and the ball in the game of billiards. 
After studying the collision process in detail, we write the (rotational) velocities of the ball and the cue after the collision.  
We also find the squirt angle of the ball for an oblique collision which represents the deviation of the ball from the intended direction. 
\end{abstract}

\maketitle


\section{Introduction } \label{sec:intro}

In 1835, Coriolis~\cite{Coriolis} studied the mathematical analysis of billiards for the first time.
To my knowledge, it takes more than one century for another work for the billiard games based on physics~\cite{Moore}.   
In Refs.~\cite{Walllace88,Crown}, the authors considered the effect of friction on collisions of billiard balls in 2-dimensions.
In Ref.~\cite{Han2005}, Han studied various aspects of the billiard physics.
Mathavan used a fast camera to analyze the ball motions in the billiard games~\cite{Mathavan1} and studied the motion of the ball under cushion impact~\cite{Mathavan2010}.
The billiard models were also used to develop a robotics system~\cite{Nierhoff}.

A cue stroke on a billiard ball makes the ball run along the line of cue incidence. 
This ordinary point of view in billiards games is correct for a head-on collision only between the ball and the cue.
Coriolis~\cite{Coriolis} calculated the outcome of the collision based on a coefficient of restitution (COR) when the cue strikes the ball's center with zero impact parameter.
Here, the COR is defined based on the head-on collision.
The impact parameter $b$ represents the distance from the ball's center to the incident line of the cue.
For collisions with a non-vanishing impact parameter, he had tried the following ansatz: ``the fraction of kinetic energy loss is the same as that without the impact parameter."
This ansatz must be refined by using the impact dynamics now a days. 

In the presence of an impact parameter $b \neq 0$, the ball can be given top/back-spin or side-spin by striking it above/below or left/right side of its center. 
Moore~\cite{Moore} presented part of the analysis on the stroke with impact parameter. 
To avoid miscue, the phenomenon that the cue slides on the ball so that the ball deviates highly from the intended direction, the impact parameter should satisfy~\cite{Alciatore TP21}\footnote{We refine this formula later in this work.}  
$$
\mu_{\rm static} \geq \tan \phi = \frac{b/R}{\sqrt{1-(b/R)^2}} ,
$$
where $R$ and $\mu_{\rm static}$ denote the radius of the ball and the coefficient of static friction between the cue-tip and the ball, respectively. 
When a cue strikes above/below from its center, the ball may run along the intended path with top/back-spin.

However, when a cue strikes a side of the center, the ball heads off along a path parallel to the resultant of the normal and the friction forces. 
In this oblique collision, a situation commonly encountered in billiards, the ball does not follow the line of incidence but deflects from the direction by a few degrees.
This deflection is commonly known by billiards players~\cite{colostate,Shepard} as ``squirt'' because the ball squirts away from its intended path.
Anecdotal evidence indicates that one can reduce the squirt angle by using a light cue-tip.
In Ref.~\cite{Cross2008}, Cross presented experimental data for the squirt angle.
He also presented part of the theoretical resolution based on the bouncing ball model~\cite{Cross2002}.

However, we fail to find a literature which deals a complete theoretical analysis on the squirt based on the impact dynamics which considers the effect of friction during the collision.
Naturally, a mathematical description on the resulting motions of the ball after the impact based on the fundamental principle of physics, which are one of the most important parts of the billiard play, is absent. 
In this work, we analyze the collision process between a cue and a ball based on the impact dynamics, and analyze the phenomena happening in the process. 
We find the final state of the ball after the collision including the squirt, the central velocities, and the angular velocities of the ball and the cue.

\section{Collision between a billiard ball and a rigid cue-stick} \label{sec:4.1}

\begin{figure}[htb]
\begin{center}
\begin{tabular}{c}
 \includegraphics[width=0.4\linewidth,origin=tl]{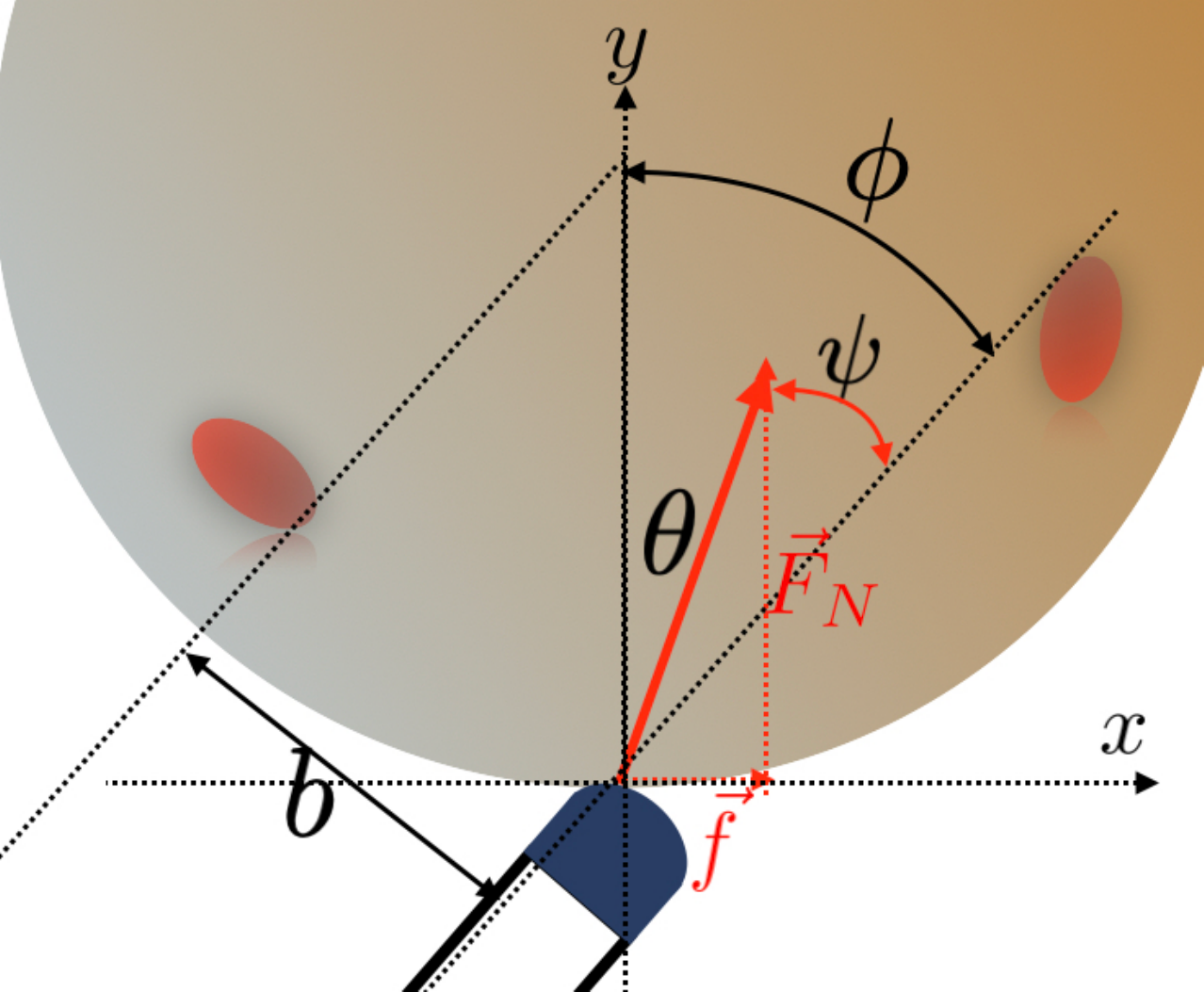} 
\end{tabular}
\end{center}
\caption{The collision between the cue and the ball. 
}
\label{fig:impact}
\end{figure}

In figure~\ref{fig:impact}, a cue stick approaches a billiard ball of radius $R$ with impact parameter $b$.
The radius of the cue-tip is about $6$ mm.
So the contact point of the cue-tip does not generally coincide with the central axis of the cue. 
The normal force $\vec{F}_N$ acts along a line from the contact point to the center of the ball, while the friction force $\vec{f}$ acts at right angles to $\vec{F}_N$.
The resultant force $\vec{F} = \vec{F}_N + \vec{f}$ acts at an angle $\theta$ to the normal-direction, where $\tan \theta = F_N/N$ defines an effective coefficient of friction between the cue-tip and the ball. 
If the tip slides on the ball, $\tan \theta = \mu_k$, where $\mu_k$ is the coefficient of sliding friction. 
If the tip grips the ball, then $\mu < \mu_k$. 
If the ball initially static, it will exit from the cue along the direction parallel to $\vec{F}$ at an angle $\psi$ from the line of incidence of the cue. 
The angle $\psi$ commonly called the squirt angle describes the undesirable deflection of the ball from the intended path. 

The experimental data in Ref.~\cite{Cross2008}, for well-chalked tips, show that the squirt angle increases almost linearly for impact parameter $b < 0.5$ and shows a bit of non-linearlity for larger impact parameters $b\sim 0.5 R$.  
The sidespin obtained from the stroke increases almost linearly with the impact parameter. 
Even though the experiment was performed with an unusual cue that produces large squirt angles, the qualitative characteristics will be the same as those of the ordinary cue.  

Let us analyze the physical situation in detail. 
We choose the coordinate system so that we can deal the collision between the ball and the cue-stick easily. 
If we stroke the cue-stick horizontally, we can ignore the friction between the ball and the table, which does not play any role.
Then, the only important thing is the relative position between the cue-stick and the ball. 
In this case, without loss of generality, we locate the coordinate's origin at the impact point between the ball and the cue.
Besides, the ball's surface at the contact point is parallel to the $(x,z)$ plane as in Fig.~\ref{fig:impact}. 
The direction to the center of the ball is now the $y$-direction and the cue-stick moves on the $(x,y)$-plane with the angle $\phi$ with respect to the $y$-coordinate in clock-wise direction. 

We assume the cue-stick is a rigid body and analyze the collision with the ball in this section. 
However, an ordinary cue-stick consists of wood, flexible to impacts orthogonal to the symmetry axis. 
We consider the effect of this flexibility in the continuing sections. 
In this work, we assume that we hold the cue very lightly so that the hand's force does not affect the ball during the impact.

Let us call the cue-part of the impact point $C_{\rm cue}$ and the ball-part $C_{\rm ball}$.
Then, the displacement vector from the ball's center to the impact point is
\be{Ri}
 \vec{R}_I  \equiv (X_I, Y_I, 0)=  (0, -R,0) = - R \,\hat y .
\ee
The displacement vector from the center of mass of the cue to the impact point is, along the incidence line of the cue, 
\be{rc}
\vec{r}_c  = (x_c, y_c, 0) =  r_c (\sin \phi, \cos \phi  , 0), \qquad 0 \leq \phi < 45^{\rm o} .
\ee
In a usual billiard games with the friction coefficient $\mu \sim 0.7 $, to avoid cue-miss, the impact point should satisfy $\phi < 35^{\rm o}$ usually. 
In this work, we choose the upper bound of $\phi$ to be $\phi= 45^{\rm o}$ because it is almost the same as the Coriolis' limit of the impact parameter $b_{\rm max}=0.7 R < R \sin 45^{\rm o} \approx 0.707 R$.

If the cue is a rigid body, $r_c$ corresponds to the distance from the the cue-tip to the center of mass of the cue, which is 0.7 times the length of the cue roughly. 
However, for an impact which acts instantly, the cue does not react as if it is a rigid body. 
In this section, we regard the cue acts as if it is a rigid body and formulate the collision process in general.
The non-rigidity which affects on the squirt angle crucially will be dealt in Sec.~\ref{sec:4.4}.
In this section, we study the details which happens during the collision. 

Now, let us write the equation of motions for the cue and the ball. 
Let the masses of the cue and the ball be $M_c$, $M$ and their velocities be $\vec{U}$ and $\vec{V}$.
Let us write their angular velocities as $\vec{\omega}$ and $\vec{\Omega}$, respectively. 
Then, the equation of motions for of the center of masses for the cue and the ball are
\be{Newton2}
M_c \frac{d\vec{U}}{dt} = -\vec{F}, \qquad M \frac{d\vec{V}}{dt} = \vec{F} .
\ee
Here $\vec{F} = (F_\parallel, F_\perp, 0)$ represents the force of the cue acting on the ball.
Because the cue moves on the $(x,y)$-plane, there is no force along the $z$-direction.
Here, the subscript ${}_\parallel$ and ${}_\perp$ represent the parallel $x$-direction and orthogonal $y$-direction to the impact plane, respectively. 

The force exerted by the cue on the ball has two origins: 
One is the normal force $F_\perp$ to the ball's surface related to the cue-tip elasticity and the other is the friction force $F_\parallel$ parallel to the surface. 
So the force $\vec{F}$ is generally not parallel to the incidence line of the cue.
Adding and integrating the two equations above, we get the following law of conservation of momentum:
\be{P:conserv}
M_c \vec{U} + M \vec{V} = M_c \vec{\vi},
\ee
where the right-hand side denotes the initial momentum of the cue.

Let the rotational inertia of the cue and the ball be $I_c= \nu_c M_c r_c^2$ and $I = \nu M R^2$, respectively.
For the case of a ball of uniform density $\nu= 2/5$ and for an object with a skewed mass such as a cue, it is roughly $\nu_c \simeq 0.136$, respectively. 
Then, the angular equation of motion of the cue is 
\be{Rot}
I_c \frac{d\vec{\omega}}{dt} = \vec{r}_c \times (-\vec{F}) = r_c( \cos \phi F_\parallel - \sin\phi F_\perp)\hat z .
\ee
Here,  $\vec{r}_c$ is the displacement~\eqref{rc} from the center of mass of the cue to the impact point $C_{\rm cue}$.
The rotational equation of motion of the ball is 
\be{Rot:ball}
I\frac{d\vec{\Omega}}{dt}  = \vec{R}_I\times \vec{F} = R F_\parallel \, \hat z .
\ee
As you can see from this equation, both the cue and the ball rotate around their own $z$-axis.

To deal the collision between the cue and the ball, we need to consider the relative motion of the two impact points.
The velocity of the impact point of the cue, $C_{\rm cue}$, is $\vec{U} + \vec{\omega} \times \vec{r}_c $ and the velocity of the impact point of the ball, $C_{\rm ball}$, is 
$
\vec{V} + \vec{\Omega}\times \vec{R}_I .
$
The relative velocity $\vec{v}$ between the impact point of the ball and that of the cue-tip can be obtained by subtracting the former from the latter: 
\be{vrel}
\vec{v} \equiv (v_\parallel, v_\perp, 0)
= \vec{V}-\vec{U}-\vec{R}_I \times \vec{\Omega}  + \vec{r}_c\times \vec{\omega} .
\ee
Conversely, the relative velocity of the cue-tip $(C_{\rm cue})$ with respect to the impact point of the ball is $-\vec{v}$ and the cue-tip will be contracted by the reaction force $-\vec{F}$ of the ball to the cue-tip.  
Because the cue-tip is more resilient than the ball, the elastic phenomena happening during the collision are mainly due to the cue-tip.  
 
The change of the relative velocity, applying the equation of motions~\eqref{Newton2}, \eqref{Rot}, and \eqref{Rot:ball}, is described by 
\bea
\frac{d\vec{v}}{dt} &=&  -\frac{d\vec{U}}{dt} +\frac{d\vec{V}}{dt} +\vec{r}_c\times \frac{d\vec{\omega}}{dt} - \vec{R}_I \times \frac{d\vec{\Omega}}{dt}  \nn \\
&=& \frac{1}{m} \vec{F} -\frac{\vec{r}_c \times( \vec{r}_c \times \vec{F}) }{I_c}  
	- \frac{\vec{R}_I \times(\vec{R}_I \times \vec{F})}{I} . \nn
\eea
Here, the mass parameter $m$ is
\be{m:Mc M}
m \equiv \frac{M M_c}{M + M_c}.
\ee
Using the vector relation
$
\vec{r}_c \times( \vec{r}_c \times \vec{F})= - r_c^2  \vec{F} + (\vec{r}_c\cdot \vec{F}) \vec{r}_c,
$ 
we get
\be{dv:FF}
\frac{d\vec{v}}{dt} =
	\frac{1}{m}\left[1 +\frac{m r_c^2 }{I_c} + \frac{mR_I^2}{I}\right] \vec{F}
- \frac{(\vec{r}_c\cdot \vec{F}) \vec{r}_c}{I_c} - \frac{(\vec{R}_I \cdot \vec{F}) \vec{R}_I}{I} .
\ee 

Let us write the amount of impulse that the cue acts on the ball for a short period $dt$ along the direction parallel to and perpendicular to the plane of impact to be
\be{impulse}
dp_\parallel \equiv F_\parallel dt = M dV_\parallel , \qquad 
  dp_\perp \equiv F_\perp dt =M dV_\perp ,
\ee
respectively.
Then, we write the equation of motion~\eqref{dv:FF} for the relative velocity by means of the impulses:
\be{dvdv}
dv_\parallel = \frac{\beta_\parallel dp_\parallel - \beta_\times dp_\perp}{m} ,  \qquad
dv_\perp = \frac{- \beta_\times dp_\parallel+\beta_\perp dp_\perp }{m} . 
\ee
In describing the collision process, we follow the construction of Stronge~\cite{Stronge2018}.
For notational convenience, we have introduced three constants,
\bea \label{beta:r}
\beta_\times &=& \frac{\sin\phi \cos \phi}{\nu_c(Q_0+1) }, \qquad  \nn\\
\beta_\parallel &=& 1 + \frac{Q_0}{\nu (Q_0+1)}+ \frac{\cos^2\phi}{\nu_c(Q_0+1)}, \qquad \nn \\
\beta_\perp &=& 1+\frac{\sin^2\phi}{\nu_c(Q_0+1)} .
\eea
Here,  $Q_0 = M_c/M$ and $\nu_c \approx 0.136$. 
These constants satisfy
\be{gamma:beta}
\beta_\parallel, \beta_\perp > 0, \qquad 
0\leq \gamma \leq 1; \qquad 
\gamma \equiv \frac{\beta_\times^2}{\beta_\parallel \beta_\perp }.
\ee
Because the friction force does not change its direction on the $(x,z)$-plane and the constant $\beta_k$ does not change, one can sum Eq.~\eqref{beta:r} over the changes of the impacts $dp_\perp$ and $dp_\parallel$. 

The impact $p\equiv p_\perp$ orthogonal to the impact plane increases monotonically during the collision where $\vec{F} \neq 0$.
Therefore, we describe the collision process by using this parameter instead of time. 
We omit the subscript ${}_\perp$ because $p$ will be used widely here. 

Let us write the initial values. 
The initial orthogonal impact $p$ must be zero because there is no impact before the cue-tip touches the ball. 
When the cue touches the ball, the velocities of the cue normal and parallel to the impact plane are $\vi \cos \phi$ and $\vi \sin\phi$, respectively.
Before the collision, the ball is static. 
Therefore, the initial value of the relative velocity $\vec{v}$ is 
\be{v0}
v_\perp (0)=v_{\perp 0}=-\vi\cos \phi, \qquad 
v_\parallel (0) =v_{\parallel 0}= -\vi \sin \phi .
\ee

The velocity and the angular velocity change according to the equations~\eqref{Newton2} and \eqref{Rot}, respectively. 
The changes of the center of mass velocities of the cue and the ball  satisfy
\be{dU, dV}
M_c dU_\perp = - dp_\perp ,\quad M_c dU_\parallel = -d p_\parallel,  \quad 
M dV_\perp = dp_\perp, \quad M dV_{\parallel} = dp_\parallel . 
\ee
The angular velocities of the cue and the ball, according to the equation, follow equation, 
\be{domega dOmega}
I_c d\omega_z = r_c (\cos \phi \, dp_\parallel - \sin \phi \, dp_\perp) , \qquad 
I d\Omega_z  = R dp_\parallel .
\ee
Because the mass, the rotational inertia, the angle $\phi$ are independent of time, we can obtain the velocities once we know the total impact after the collision. 

\section{The collision process between the ball and the cue} 
\label{sec:4.2}

When we strike a ball off the center to give the ball a spin, first of all,   the cue-tip starts to slide on the ball's surface.
That is because the sliding velocity $\vi \sin \phi$ cannot be made to zero instantly unless the friction force may not be indefinitely large. 
Because of the normal force acting on the cue-tip, elastic energy accumulates as the tip contracts.
The parallel friction force between the tip and the ball reduces the sliding velocity. 
The slip disappears at a moment while the cue-tip is contracting due to the frictional force.
Later, the tip grips the ball and moves together.
As we know, the maximum static frictional force always equals to or is greater than the kinetic frictional force.
Therefore, once the slip disappears, an additional slip will not happen.
At some point, the cue-tip contracts to its maximum, and the elastic energy accumulated until then will restore the shape of the tip and pushes the ball out.
By the end of the contact period, the friction force is no longer enough to hold the ball firmly, so the ball bounces off the compressed tip.

In general, it is unclear which is first between the moment when the maximum compression happens and the moment when the slip disappears.
The order must be dependent on the initial values.
In general collisions, we analyze the maximum compression first case separately from the slip disappearance first case. 

In the case of the billiards, as we will see in \ref{sec:4.4}, the slip disappears almost instantaneously when the cue touches the billiard ball.
So in this work, we consider only the formal.

\vspace{.1cm}
Let us explain the collision process in the order of time. 
\begin{enumerate}
\item 
If the impact parameter $b= R\sin\phi \neq 0$ when the cue-tip touches the ball, it enters the {\it slip state}.
The cue collides with the ball with a vertical velocity of $\vi\cos \phi$.
At the same time it slides on the surface of the ball with a velocity $\vi \sin \phi$.
Because the friction force is finite, the slip of the cue never stops instantaneously.
The existence of the slip state is the origin of the squirt when striking off the center of the ball.

If the kinetic friction coefficient is $\mu$, the amount of friction force $f$ on the ball is proportional to the magnitude of the normal force ($= F_\perp$).
 So the amount of impulse parallel to the impact surface by the frictional force satisfies
\be{friction:p}
dp_\parallel = \mu \, dp_\perp .
\ee
Here, the kinetic friction coefficient is assumed to be independent of the speed at which the cue slides and the vertical pressure, which holds when the sliding is not too fast on most surfaces.
If the friction coefficient is large enough and the cue-miss does not happen, the sliding state will stop soon.
Then, the cue and the ball will change to the `stick state'.
Let the amount of the vertical impact at the moment of this change be $p = p_s$.
In other words, we mark {\it the moment when slip stops and changes to `stick state'} as $p_s$.
  
During the slip, from the equation ~\eqref{dvdv}, the horizontal and the vertical components of the relative velocity are
\be{sliding}
\frac{dv_\parallel}{dp_\perp} = \frac{\mu \beta_\parallel 
	- \beta_\times}{m} ,  \qquad
\frac{dv_\perp}{dp_\perp} = \frac{- \mu\beta_\times +\beta_\perp }{m} ,  \quad 0 \leq p_\perp < p_s .
\ee
The initial value of $v_\parallel$ is negative, because of Eq.~\eqref{v0}.
If we want the sliding speed $|v_\parallel|$ to decrease, the kinetic friction must be large enough to satisfy $dv_\parallel > 0$, i.e.,
\be{mu:cond1}
\mu \geq \mu_c \equiv \frac{\beta_\times}{\beta_\parallel}.
\ee

In ordinary collisions, if the friction coefficient is not large enough, the slip can reverse its direction due to the frictional force. 
But there is no such phenomenon in the collision between the cue and a billiard ball if
the friction coefficient is large enough to stop slip so that a cue-miss has not occurred.
Then, the cue grabs the ball soon, i.e., $v_\parallel (p_s)=0$.
Integrating Eq.~\eqref{sliding}, the relative velocity at $p= p_s$ becomes  
\be{v:v0}
0= v_\parallel(p_s) = v_{\parallel 0} + \left(\frac{\mu}{ \mu_c}-1\right) \frac{\beta_\times p_s}{m}, \qquad
v_\perp(p_s) = v_{\perp 0} +\left(1- \frac{\mu \gamma}{\mu_c} \right) \frac{  \beta_\perp p_s}{m} .
\ee 
At $p= p_s$, we have $v_\parallel =0$ and $dv_\parallel =0$. 
Then, we can write the vertical impulse $p_s$ in terms of the the initial values and the kinetic friction coefficient:
\be{ps:v}
p_s = \frac{-m v_{\parallel0} }{ \beta_\times (\mu /\mu_c -1)}. 
\ee
The vertical part of the relative velocity at this moment is 
\be{ps:v5}
v_\perp(p_s) = v_{\perp0}\left[1- \frac{1- \gamma \mu/\mu_c  }
	{\mu/\mu_c-1}
	 \frac{ \mu_c}{\gamma } \tan \phi \right] .
\ee
Here, its initial value is $v_{\perp 0} = -\vi \cos \phi$.
The constants $\gamma$ and $\mu_c$ are determined from Eqs.~\eqref{gamma:beta} and~\eqref{mu:cond1}. 

The amount of the horizontal impulse, $p_\parallel$, delivered to the ball during the slip state, by using Eq.~\eqref{friction:p}, is
\be{p parallel ps}
p_\parallel(p_s) = \mu p_s = \frac{\mu m \vi \sin \phi }{ \beta_\times (\mu /\mu_c -1)}.
\ee
As expected, this amount is proportional to the cue's horizontal velocity.
It vanishes if the friction coefficient vanishes.
In that case, the ball gets only the vertical velocity and will proceed in the $y$-direction, which is very different from the direction of the cue.

The vertical relative speed may increase or decrease depending on the initial condition.
Since we expect the ball to separate over time, we generally expect the vertical relative velocity to increase continuously, then satisfies
\be{cond:pp}
 dv_\perp > 0 \, \rightarrow \, \mu_c > \gamma \mu .
\ee
Combining this with Eq.~\eqref{mu:cond1}, we obtain a constraint on the kinetic friction coefficient $\mu$:
\be{mu:cond}
1  \leq  \frac{\mu}{\mu_c} < \gamma^{-1}
\ee

Under specific circumstances, this vertical relative speed may initially decrease. 
In this case, it is said that a 'jam' occurred.
When the `jam' occurs, $\gamma> \mu_c/\mu$ holds.
At the moment the slip stops, the 'jam' also ends.
Under ordinary circumstances, if the sliding state ends during the compression of the cue-tip, $dv_\parallel =0$ and $v_{\perp0}<v_\perp(p_s) <0 $ must be satisfied at $p =p_s$. 
That is, the cue-tip is still undergoing compression at the time.
Then, from Eqs.~\eqref{ps:v} and~\eqref{ps:v5}, the initial value must satisfy 
\be{cond:stick}
\frac{v_{\parallel0}}{v_{\perp0}} 
= \tan \phi \leq \frac{1}{\mu_c}\frac{\mu/\mu_c-1} {1/\gamma-  \mu/\mu_c  }
\ee
if we want the sliding state stops during the compression undergoes. 
If this equation is not satisfied, the cue does not go into the `stick state' during the compression process, but makes a cue-miss or becomes stuck during the restoration process.
Conversely, this equation determines the lower bound for $\gamma$:
\be{lower}
\gamma \geq \left( \frac{\mu/\mu_c -1}{\mu_c \tan \phi } + \frac{\mu}{\mu_c} \right)^{-1} .
\ee

Next, we find the energy used to compress the cue-tip while the cue slides on the ball.
Suppose that the particle is moved by the displacement $x$ along the direction of the force $\vec{F}$ from $t=0$ to $t=t_f$.
Then, the work done by the force during this time is
$$
W_f = \int_0^x F dx' = \int_0^{t_f} F v dt = \int_0^{p_f} v dp , \qquad dp = F dt .
$$
If the normal velocity changes linearly with respect to the vertical impulse $p$, the work done by the force during $p_i< p< p_j$ is 
\be{W:p}
W = \int_{p_i}^{p_j} v dp = \frac{v_j +v_i}{2} (p_j-p_i) 
\ee
where $v_i$ and $v_j$ are the corresponding values of normal velocity. 

From this equation, the energy absorbed by the cue-tip compression during the slip state, $0 \leq p_\perp \leq p_s$, is
\be{W perp ps}
W_\perp (p_s) = \frac{(v_\perp(p_s) + v_{\perp0}  ) p_s}{2} = v_{\perp0} p_s +
	\left(1-\frac{\gamma \mu}{\mu_c}\right) \frac{\beta_\perp p_s^2}{2m} .
\ee
 The energy absorbed in this way changes the shape of the cue-tip near the impact point and is saved as deformation energy of the cue-tip. 
 At the later stage of the collision, the stored energy will be released into the forms of kinetic energy of the ball, heat, and sound.

\item After the slip has stopped $p> p_s$, the stick state begins.
Now, the impact point of the ball moves at the same speed as that of the cue horizontally.
The cue-tip and the ball will stick together due to friction.
Since the ball and cue tips are attached, there is no relative horizontal speed.
Therefore, from the expression~\eqref{dvdv}, we get
\be{stick:eom}
dv_{\parallel} = 0 \quad \Rightarrow \quad dp_\parallel = \mu_c dp, \qquad p > p_s .
\ee

From Eq.~\eqref{dvdv}, the change of the normal relative velocity in the absence of slip is
\be{dv+}
dv_\perp = (1-\gamma)\frac{\beta_\perp \, dp}{m},
 \qquad  p > p_s .
\ee
Also, the horizontal impulse cumulated during this stick state is, by integrating Eq.~\eqref{stick:eom},
\be{p II:stick}
p_\parallel (p)= \mu p_s + \mu_c (p - p_s), \qquad p> p_s .
\ee

\item 
Because of the vertical velocity, the cue-tip continues to compress, accumulating energy.
At the moment of maximal compression, the vertical compression stops $ v_\perp = 0$.
Let the amount of impulse at the moment be $p = p_c$. 
Then, $p_s< p_c$ because vertical compression stops when the tip and the ball are stuck.
Integrating the expression in Eq.~\eqref{dv+} from $p=p_s$ to $p=p_c$ and setting $p=p_c$ one gets
$$
0 = v_\perp (p_c) = v_\perp (p_s) + (1-\gamma) \frac{\beta_\perp(p_c- p_s)}{ m} .
$$

Of course, the relative speed in the horizontal direction vanishes, $v_\parallel(p_c)=0$, because they are stuck.
During this additional compression period, from the above equation, an additional vertical impulse 
\be{pc}
p_c -p_s= - \frac{m v_\perp(p_s)}{(1-\gamma) \beta_\perp} 
\ee
acts to the ball.

Therefore, the amount of impulse up to the moment of maximum compression, in terms of the initial values using the equation ~\eqref{ps:v}, is
\be{pc2}
p_c =\frac{m (-v_{\perp0})}{(1-\gamma) \beta_\perp} 
	\left[1+ \mu_c \tan \phi \right].
\ee 
Here, $\tan \phi = v_{\parallel 0}/v_{\perp 0}$.
For later convenience, we write $v_{\perp 0}$ in terms of $p_s/p_c$,
\bea
v_{\perp 0} = \frac{- p_c (1-\gamma) \beta_\perp}{m(1+ \mu_c \tan\phi)}
&=& \frac{(1-\gamma) \beta_\perp p_c}{m} \Big( \frac{\mu_c \tan \phi}{1+\mu_c \tan\phi} -1\Big) \nn \\
&=&  \frac{(1-\gamma) \beta_\perp p_c}{m} 
\Big(\frac{\gamma(\mu/\mu_c-1)}{1-\gamma} \frac{p_s}{p_c} -1\Big). 
	\label{v perp 0}
\eea

By using Eq.~\eqref{W:p}, we get the energy cumulated by the normal force during the stick state, because $v_\perp(p_c)=0$, to be 
\be{Wpc}
W_\perp (p_c) - W_\perp(p_s) = \frac{v_\perp(p_c)+ v_\perp(p_s)}{2} (p_c - p_s) 
= -\frac{\beta_\perp (1-\gamma)}{2m}(p_c - p_s)^2 .
\ee
Summing this with the energy in Eq.~\eqref{W perp ps}, the total energy absorbed during the whole compression process including the `slip state' is
\bea
W_\perp(p_c) &=& W_\perp(p_s)  + [W_\perp (p_c) - W_\perp(p_s) ]\nn \\
&=& v_{\perp0} p_s +
	\Big(1 - \frac{\gamma \mu}{\mu_c} \Big) \frac{\beta_\perp \,p_s^2}{2m} 
	-(1-\gamma)\frac{\beta_\perp(p_c - p_s)^2 }{2m}\nn \\
&=& -\frac{(1-\gamma)\beta_\perp  p_c^2}{2m} 
\left[  1- \frac{(\mu/\mu_c-1) \gamma}{1-\gamma} \left(\frac{p_s}{p_c}\right)^2 
	\right]     .          \label{W pc}
\eea
Here, we use Eq.~\eqref{v perp 0} and
\be{ps/pc}
\frac{p_s}{p_c} = \frac{(1-\gamma)\mu_c \tan \phi}{\gamma (\mu/\mu_c -1)( 1+\mu_c\tan \phi )}  .
\ee

\item 
As soon as the vertical velocity of the cue becomes the same as that of the ball, the compression stops, and the {\it restoration process} begins.
During this process, the shape of the cue-tip returns to its original form, and the elastic energy stored during the compression process is released to push the ball.
Let's call the terminal value of the vertical impulse obtained in the restoration process as $p_f$ and find the restored kinetic energy.

Since vertical compression stops while the cue-tip and the ball are attached, $p_s<p_c$ is satisfied.
In this case, the tip and the ball remain attached until the ball is finally separated at $p= p_f$.
We integrate the equation~\eqref{dv+} for $p_c\leq p \leq p_f$ to get the terminal vertical relative velocity,
\be{vf:pf}
v_\perp(p_f) = \frac{(1-\gamma) \beta_\perp}{m} ( p_f - p_c) .
\ee
The terminal relative velocity parallel to the impact plane is $v_\parallel (p_f)=0$ because it is in the stick state.

The energy recovered in the restoration process from the elasticity of the cue-tip, by using Eq.~\eqref{W:p}, is 
\be{W:rest}
W_\perp(p_f) - W_\perp(p_c) =
\frac{v_\perp(p_f) + v_\perp(p_c)}{2} ( p_f-p_c) 
=  \frac{(1-\gamma) \beta_\perp p_c^2}{2m} 
	\Big( \frac{p_f}{p_c} - 1\Big) ^2 .
\ee

Also, the impulse in the horizontal direction cumulated from the initial slip state to the final restoration period is calculated from Eq.~\eqref{p II:stick}:
\be{p II f}
p_\parallel(p_f) = \mu_c p_c \left[\Big(\frac{\mu}{ \mu_c}-1\Big)\frac{p_s}{p_c} 
			+  \frac{p_f}{p_c}\right] .
\ee
\end{enumerate}

\vspace{.3cm}
So far, we have looked into the collision process of the cue and the ball in detail.
In describing the process, we follow the construction of Stronge~\cite{Stronge2018}.
He defined the (energetic) coefficient of restitution (ECOR) from the ratio of the elastic energy absorbed in the compression process to that recovered in the restoration process as
\be{e*}
e_*^2 = - \frac{W_\perp (p_f) - W_\perp(p_c)}{W_\perp(p_c)} .
\ee
Therefore, the ECOR, by using Eq.~\eqref{W pc} and \eqref{W:rest}, is 
\be{e* 1}
e_*^2 =\frac{ ( \frac{p_f}{p_c} - 1) ^2 
}{ 1- \frac{(\mu/\mu_c-1) \gamma}{1-\gamma} \left(\frac{p_s}{p_c}\right)^2   } .
\ee
From this result, we can find the ECOR once we know the initial values and $p_f$.

Conversely, given the ECOR, the ratio $p_f/p_c$ can be found by arranging Eq.~\eqref{e* 1}: 
\be{pf/pc}
\frac{p_f}{p_c} = 1+ e_* \sqrt{ 1- \frac{(\mu/\mu_c-1) \gamma}{1-\gamma} \left(\frac{p_s}{p_c}\right)^2  }.
\ee
Here, since $p_f> p_c$, a positive radical was taken.
In addition, from Eq.~\eqref{vf:pf} for a given ECOR, the ratio of the vertical velocity is
\be{v perp e}
\frac{v_\perp(p_f)}{v_\perp(0)} = - e_* G,
\ee
where $G$ is the modification factor of the COR determined by the initial value
\bea \label{G}
G & \equiv & \left(1+ \mu_c \tan \phi \right)
	 \sqrt{1- \frac{(\mu/\mu_c-1) \gamma}{1-\gamma} \left(\frac{p_s}{p_c}\right)^2    }.
\eea
The impulse $p_s = 0$ and $v_{\parallel0} = 0$ if the cue strikes the center of the ball, which gives exactly the same result as the existing Newtonian COR.
However, if the impact parameter does not vanish and the cue does not strike the center of the ball, the ECOR is different from the existing COR by the proportionality factpr $G$.

\section{The state of the cue and ball after the impact} \label{sec:4.3}

In this section, we determine the vertical relative velocity after the collision using a correctly defined ECOR.

First, let's put together several equations to find the relative speed after the impact.
The initial value of the relative velocity, as shown in Eq.~\eqref{v0}, is
$
v_\parallel(0) =-\vi \sin \phi, \, ~ v_\perp(0) =-\vi \cos \phi .
$
We find the ratio of impulses at each moments by using the expressions~\eqref{ps/pc}, \eqref{G} and \eqref{pf/pc}, to be
\be{pfpc2}
\frac{p_s}{p_c} = \frac{(1-\gamma)\mu_c \tan \phi}
		{\gamma (\mu/\mu_c -1)(1+ \mu_c  \tan \phi)}  , \quad
\frac{p_f}{p_c} = 1+ \frac{e_* G}{1+ \mu_c \tan \phi} .
\ee
In addition, the impulse $p_c$ at the moment of maximal contraction is
\be{pc:ui}
p_c = p_0 ( 1+ \mu_c \tan \phi) ; \qquad 
p_0 \equiv \frac{m U_i \cos \phi}{(1-\gamma) \beta_\perp} .
\ee

Putting these values to Eq.~\eqref{G}, we get 
\be{G:res}
G = \sqrt{ (1+\mu_c \tan \phi)^2 - \frac{\gamma^{-1}-1 }{\mu/\mu_c -1} (\mu_c \tan \phi)^2 } .
\ee
Here
$
\mu_c = {\beta_\times/\beta_\parallel}, \,~
\gamma = (\beta_\times^2)/\beta_\perp \beta_\parallel = (\beta_\parallel/\beta_\perp)\mu_c^2 .
$
In addition, a bound for $\gamma$ is given from Eq.~\eqref{lower}:
\be{bound1}
\left( \frac{\mu/\mu_c -1}{\mu_c \tan \phi } + \frac{\mu}{\mu_c} \right)^{-1}  \leq \gamma \leq 1.
\ee
If a `jam' state does not happen, the upper limit of $\gamma$ is reduced to $\mu_c/\mu\, (<1)$.
However, the upper limit of $\gamma$ is 1 because a `jam' may occur in some situations.

As you can see in the expression~\eqref{G:res}, the value of $G$ behaves monotonically with respect to $\gamma$.
Then, the range of $G$ for the bound~\eqref{bound1} becomes 
$$
\left[1-\frac{\sin^2\phi}
	{(\nu+1)(Q_0 +1)   }\right]^{-1/2}  \leq   G \leq  \left[1-\frac{\sin^2\phi}
	{(\nu+1)(Q_0 +1)   }\right]^{-1}.
$$
Therefore, the modification factor $G$ is always greater than 1.
In addition, both the upper and the lower bounds increase monotonically with $\phi$.

Given $G$, we determine the terminal relative velocity after the impact from their initial values:  
\be{vf:egv}
v_\perp(p_f) = - e_* G v_{\perp0} = e_* G U_i \cos \phi, \qquad
v_\parallel(p_f) =0 .
\ee

To obtain the velocities of the cue and the ball, we calculate the terminal impulse parallel to the impact plane. 
First, we put Eqs.~\eqref{ps:v}, \eqref{pc}, \eqref{ps/pc}, \eqref{pf/pc}, \eqref{G} to Eq.~\eqref{p II f}. 
Then, we get
\be{pf parl}
p_{\parallel f} 
=\mu_c p_0\left[1+ e_*G + \frac{ \mu_c\tan \phi}{\gamma} \right] .
\ee

We get the velocity $\vec{V}^0$ of the ball after the impact by dividing the terminal impulse the ball gets by the mass of the ball, since the ball was static before the impact.
In other words, $ V^0_{\perp} = p_{f}/M$ and  $V^0_{\parallel} = p_\parallel (p_f)  /M$.
Using Eqs.~\eqref{pfpc2}, \eqref{pc:ui}, and\eqref{pf parl}, we get 
\bea
V^0_{\perp} &=&\frac{m}M \frac{ U_i \cos \phi}{(1-\gamma) \beta_\perp} \left(1+ e_* G+ \mu_c \tan \phi \right) , \nn \\
V^0_{\parallel} &=&\frac{m}M  \frac{\vi \cos\phi}{(1-\gamma) \beta_\perp}  \mu_c \left[1+ e_*G + \frac{ \mu_c\tan \phi}{\gamma} \right] .
\eea
Now, the kinetic energy of translation of the ball is 
$$
\frac12 M (\vec{V}^0)^2= \frac{p_0^2}{2M} 
\left[ (1+ e_* G +\mu_c \tan \phi)^2 + \mu_c^2 \Big(1+ e_* G +\frac{\mu_c \tan \phi}{\gamma}\Big)^2 \right] .
$$

To find the squirt angle of the ball, we call the angle between the velocity of the ball after the collision and the $y$-axis $\theta $.
Then,
\be{theta f}
\tan \theta =\frac{V^0_{\parallel}}{V^0_{\perp}}
= \mu_c\Big(1 +
	 \frac{(\gamma^{-1}-1)\mu_c \tan \phi }{1+e_* G+ \mu_c \tan \phi } \Big) .
\ee
 \begin{figure}[htb]
\begin{center}
\begin{tabular}{cc}
 \includegraphics[width=.50\linewidth,origin=tl]{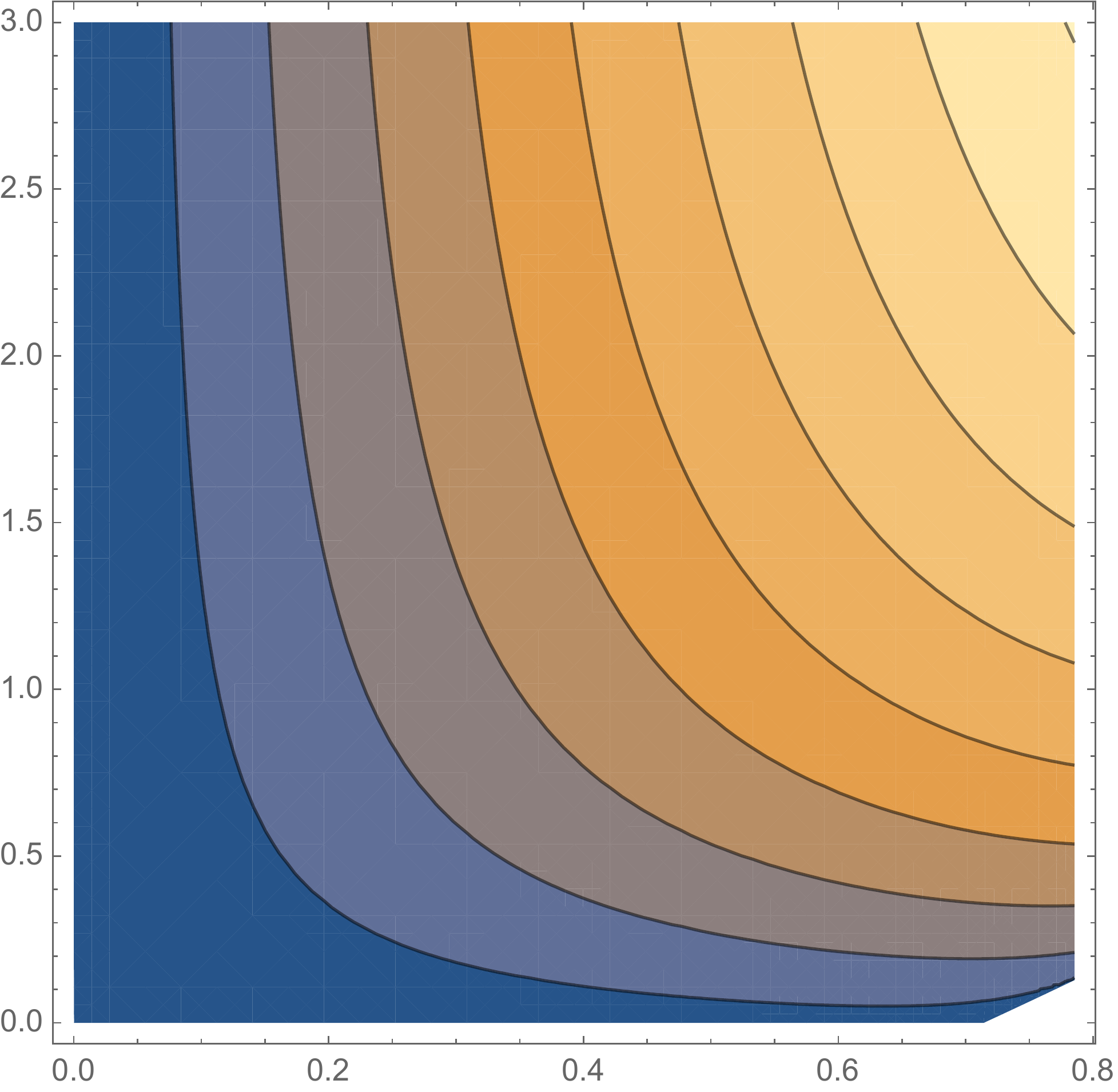} &\includegraphics[width=.06\linewidth,origin=tl]{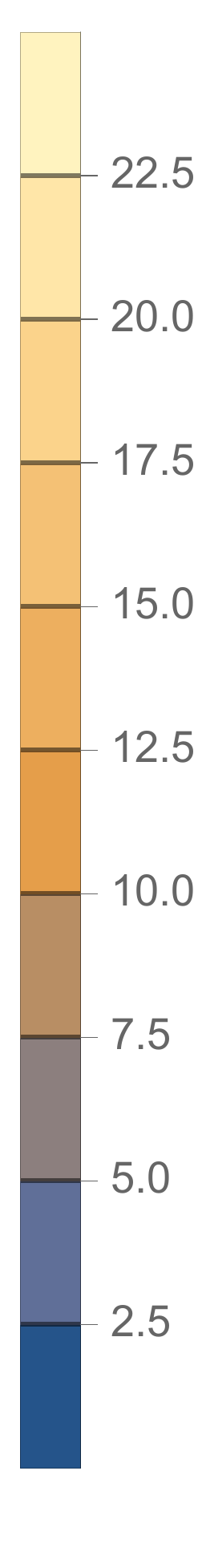}
\end{tabular}
\end{center}
\caption{The squirt angle for a rigid cue. 
The horizontal axis represents the angle $\phi$ and the vertical axis represents $ Q_0= M_c/M$. Typical values of $e_* = 3/4$ and $\mu = 0.6$ were used.}
\label{fig:squirt2}
\end{figure} 
Then, the squirt angle is 
$$
\psi = \phi - \theta .
$$
The formula for this squirt angle can be easily identified by referring to the figure~\ref{fig:cue}.
In calculating the squirt angle, we assumed that the cue is a rigid body. 
When the cue-ball mass ratio is about $M_c/M \sim 3$ and the cue strikes half the radius ($\psi =30^{\rm o} \approx 0.523$ rad.), the squirt angle is about $16^{\rm o}$.
This value is very different from the actual squirt angle that appears in billiard games.
However, we can find qualitative tendencies for the squirt angle. 
The squirt angle vanishes when $Q_0 \to 0$ and $\phi =0$ and increases with $\phi$ and $Q_0$.

Also, since the amount of squirt for a genuine cue is much smaller than this angle, we notice that the main assumption we took in this section,``the cue behaves as if it is a rigid body" is incorrect.
In the next section, we break out the assumption and see how we can build a more physically acceptable cue model.

\section{The flexibility of the cue and the impact  } \label{sec:4.4}
\begin{figure}[htb]
\begin{center}
\begin{tabular}{c}
 \includegraphics[width=.50\linewidth,origin=tl]{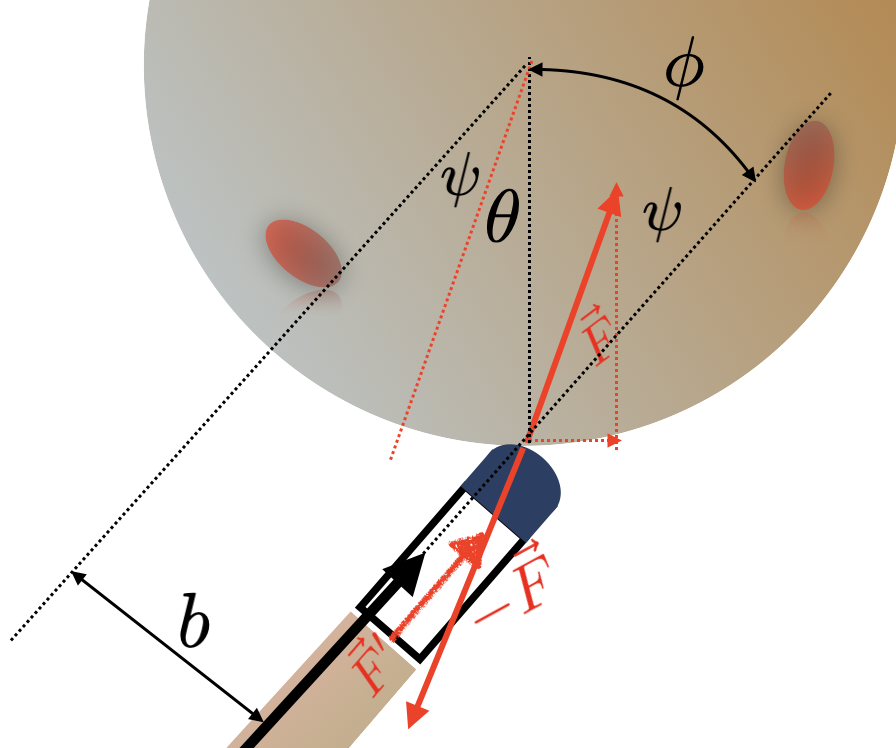} 
\end{tabular}
\end{center}
\caption{The cue model and the squirt angle $\psi$.}
\label{fig:cue}
\end{figure} 

In the previous section, we dealt with the problem by regarding the cue as a rigid body.
However, since the actual cue consists of wood, it vibrates when a perpendicular impulse to the cue line works.
Because of this, only an end part of the cue immediately reacts to an impulse acting on the cue-tip for a very short duration such as the stroke on a billiard ball.
Later, the effect of the impulse will spread out.
On the other hand, if the moment of striking the ball is much longer than the period of the vibration, it is appropriate to approximate the cue as a rigid body.

When a cue strikes a billiard ball, the moment that the cue and the ball contact each other is about 0.002 seconds.
The vibrating frequency of the end part of the cue is about $f= 250$ Hz.\footnote{I have obtained this value by using a smartphone application `spectrum' which measures the frequency of sound.}
Thus, we get a period of vibration to be about 0.004 seconds.
It takes less time than half a period to strike the billiard ball with the cue
and there is not enough time for the impact acted on the cue-tip to spread over the entire cue.
The actual functioning of the billiard cue, then, can be modeled by an end-part of the cue including the cue-tip (let's call it `the cue-end') as shown in Fig.~\ref{fig:cue}.
When trying to make a shaft from any material, one needs to pay attention to its vibration frequency to obtain better function.

If we look at the super-slow-motion video~\cite{slowmotion}, we can see the end of the cue that hit the billiard ball vibrates.
This swinging length is about ten times the diameter of the cue-tip\footnote{Of course, this is a very rough value. The length of the cue-end depends on the type of wood, the method of making the cue, etc.}.
Since the cue-tip radius is about $b=6$ mm, we estimate the cue-end length to be about 12 cm.
The radius of the opposite side of the shaft is about 10.5 mm, which is thicker than the cue-tip.
The length and the mass of the shaft are about 70 cm and 120\,g, respectively.
Therefore, the mass of the cue-end (usually called end-mass) is approximately 1/10 of the mass of the shaft when calculated using the volume ratio of the cone:
\be{cue end}
m_c \approx 12 \mbox{ g}.
\ee
Of course, this value depends on the type and the elasticity of cue.
Summarizing, the cue-end is approximately $12 $~g, has the shape of the bottom $l = 12$ cm of a cone of length $ 105.3 $ cm.
The base radius of the cone is approximately $a = 6.77$~mm.
The volume of the cue-end is $V = \frac{\pi l}{3} (a^2 + ab + b^2)$.
The center of mass is located at
$$
\bar y = \frac{\int y dV}{V} = \frac{l (a^2 + 2ab + 3 b^2)}{4 (a^2 + ab + b^2)} \approx 0.48 l
$$
from the bottom roughly.
Therefore, since the position of the center of mass is not significantly different from the center position of the cue-end (0.5 $l$), we can regard the center of mass is located at the center without a large error. 
Therefore, the rotational inertia of the cue-end, $I_c$, is
$$
I_c = \nu_c m_c r_c ^2; \qquad \nu_c \approx \frac13 .
$$
Here, $\nu_c$ is obtained from the rotational inertia of a uniform rod and $r_c \approx 0.5 \,l$.

The body (the other part of the cue than the cue-end) plays the role of acting a force (indicated as $\vec{F}'$ in the figure) on the cue-end along the line of the cue.
If necessary, you can also consider the effect of vibration of the cue-end.
Of course, because the strike ends before the cue-end performs one period of oscillation, one can treat it as a friction like Cross~\cite{Cross2008} rather than an oscillation.

We choose the coordinate system to be the same as that in the previous section.
Let the speed of the body of the cue be $\vec{V}_{\rm cue}$.
If the central velocity of the cue-end is $\vec{U}$, we can express the velocity of the cue $\vec{V}_{\rm cue}$ to be the line-of-cue part of the cue-end velocity:
\be{W:Ui}
\vec{V}_{\rm cue}  \equiv V_{\rm cue} \hat \vi = \frac{\vec{U}\cdot \vec{\vi}}{\vi^2} \vec{\vi}
	=(U_\perp \cos \phi + U_\parallel \sin \phi )  \hat{\vi} . 
\ee
Here  $V_{\rm cue}(0) =\vi$. 
Let the end-mass be $m_c$.  
Then the mass of the body part other than the cue-end is $M_c -m_c$ and the force of this body part acting on the cue-end is $\vec{F}'$.
The equation of motion for the body-part is 
\be{eom:cue-}
(M_c-m_c) \frac{d\vec{V}_{\rm cue}}{dt} =- \vec{F}' .
\ee
The equation of motion of the cue-end, because it receives  $\vec{F}'$ from the body part and gives  $\vec{F}$ to the ball, is
\be{eom4}
 m_c \frac{d \vec{U}}{dt} = \vec{F}' - \vec{F}.
\ee
The ball of mass $M$ gains force  $\vec{F}$ from the cue-end.
Therefore, the equation of motion of the ball is
\be{eom:ball}
M\frac{d\vec{V}}{dt} =\vec{F}.
\ee

In order to describe the change of the cue-end velocity $\vi$ according to the amount of impulse received by the ball, we apply Eq.~\eqref{W:Ui} to Eqs.~\eqref{eom:cue-} and \eqref{eom4} and get
\bea
\big[m_c + (M_c -m_c) \cos^2\phi \big] dU_\perp + (M_c -m_c) \cos \phi \sin \phi \, dU_\parallel =  -dp_\perp, \nn \\ 
(M_c -m_c) \cos\phi \sin\phi \, dU_\perp +\big[m_c+ (M_c -m_c) \sin^2 \phi ] \, dU_\parallel = - dp_\parallel .
\eea
Here $dp_\perp$ and $dp_\parallel$ are defined in Eq.~\eqref{impulse}.
Rewriting the equation with respect to $dU_\perp$ and $dU_\parallel$, we get 
\bea
dU_\perp &=& \frac{-\big[m_c+ (M_c -m_c) \sin^2 \phi ]dp_\perp +(M_c -m_c) \cos \phi \sin \phi \, dp_\parallel}{M_c m_c}, \nn \\
dU_\parallel &=& \frac{(M_c -m_c) \cos \phi \sin \phi \, dp_\perp-\big[m_c+ (M_c -m_c) \cos^2 \phi ]dp_\parallel }{M_c m_c} .
\eea
Using this result and Eq.~\eqref{W:Ui}, we get 
$$
dV_{\rm cue} = - \frac{\cos \phi \,dp_\perp + \sin \phi \, dp_\parallel}{M_c} .
$$
Writing the velocity change of the cue-end as a component,
\bea \label{dW}
dV_{\rm cue\perp} &=&  - \frac{\cos^2 \phi \,dp_\perp + \cos \phi\sin \phi \, dp_\parallel}{M_c}, \nn \\
dV_{\rm cue\parallel} &=&  - \frac{\sin \phi\cos \phi \,dp_\perp + \sin^2 \phi \, dp_\parallel}{M_c} .
\eea

Let us write the angular velocities of the cue-end and the ball $\vec{\omega}$ and $\vec{\Omega}$, respectively. 
Then, the rotational equation of motion for the cue-end and the ball are
\be{rot4}
I_c \frac{d\vec{\omega}}{dt} = \vec{r}_c \times (- \vec{F}), \qquad
I \frac{d\vec{\Omega}}{dt} = \vec{R}_I \times \vec{F}  .
\ee
The force $\vec{F}'$ does not contribute to the rotational motion because this force always points toward the center of the cue-end.
The rotational inertia of the cue-end is $I_c = \nu_c m_c r_c^2$.
The relative velocity of the impact points between the cue tip and the ball is also given by the result obtained in the previous section~\eqref{vrel}, and the change of this relative velocity, by using Eqs.~\eqref{eom:cue-}, \eqref{eom4}, and \eqref{eom:ball}, is
\bea
\frac{d\vec{v}_{\rm rel}}{dt} &=&  -\frac{d\vec{U}}{dt} +\frac{d\vec{V}}{dt} +\vec{r}_c\times \frac{d\vec{\omega}}{dt} - \vec{R}_I \times \frac{d\vec{\Omega}}{dt}  \nn \\
&=& 
\frac{(M_c-m_c)}{m_c} \frac{d\vec{V}_{\rm cue}}{dt}+ \frac{1}{m} \vec{F} -\frac{\vec{r}_c \times( \vec{r}_c \times \vec{F}) }{I_c}  
	- \frac{\vec{R}_I \times(\vec{R}_I \times \vec{F})}{I} . \nn
\eea
Here, the mass $m$ represent not the mass~\eqref{m:Mc M} in the previous section but is determined by the cue-end and the ball by
\be{m:M mc}
m \equiv \frac{M m_c}{M + m_c}.
\ee
Using the vector relation
$
\vec{r}_c \times( \vec{r}_c \times \vec{F})
	= - r_c^2  \vec{F} + (\vec{r}_c\cdot \vec{F}) \vec{r}_c
$  
and Eq.~\eqref{dW}, we remove $\vec{F}'$ in the equation.
Then, we write the change of the relative velocity $dv_\perp$ and $dv_\parallel$ by using the impulse~\eqref{impulse} acting on the ball during a short period of time $dt$:
\be{dvdv2}
dv_\parallel =  \frac{\beta'_\parallel dp_\parallel - \beta'_\times dp_\perp}{m} ,  \qquad
dv_\perp = \frac{- \beta'_\times dp_\parallel+\beta'_\perp dp_\perp }{m} .
\ee
Here, the constant $\beta_k'$ with $k=\parallel, \perp, \times$ are 
\bea
\beta'_\parallel &\equiv & 
1+ \frac{m}{\nu M} +\frac{m \cos^2\phi}{m_c}\left[ 
  \frac{1}{\nu_c}- \Big(1-\frac{m_c}{M_c}\Big) \tan^2\phi\right], \nn \\
\beta'_\perp &\equiv & 1  -\frac{m\cos^2\phi}{m_c} \left[1-\frac{m_c}{M_c}-  \frac{\tan^2\phi}{\nu_c}\right]  , \nn \\
\beta'_\times & \equiv & \frac{m}{m_c }\left[\frac{1}{\nu_c}
	+1 -\frac{m_c}{M_c} \right]\cos\phi \sin \phi .
\eea
The constant $\beta'_k$ can be obtained formally from $\beta_k$ defined in the previous section by adding a term considering the cue-structure and still satisfies  
\be{gamma:'}
\beta_\perp' , \beta_\parallel' \geq 0, \qquad \gamma = \frac{(\beta'_\times)^2}{\beta'_\perp \beta_\parallel'} \leq 1 .
\ee
Here $\nu_c \approx 1/3$.
In addition, because the mass of the cue-end is much smaller than that of the ball, we have $m/m_c =(1+ m_c/M)^{-1} \sim 1-m_c/M$. 

Since the friction force does not change direction in the $(x,z)$ plane and the values of $\beta'_k$ hardly change during the collision, we can find the velocity $\vec{v}$  by integrating with respect to $p_\perp$ and $p_\parallel $.
The initial value of the relative velocity of the impact point $C_{\rm ball}$ of the ball to that of the cue is given by Eq.~\eqref{v0} as before.
Now all the calculations done in the previous sections will hold with the same form by replacing them with $\beta_k \to \beta'_k$.

Therefore, after putting $\beta_k \to \beta'_k$ in the result of the previous section, we calculate $G$, the modification factor of COR, from~\eqref{G:res},
\be{G:res2}
G = \sqrt{ (1+\mu_c \tan \phi)^2 - \frac{\gamma^{-1}-1 }{\mu/\mu_c -1} (\mu_c \tan \phi)^2 } .
\ee
Of course the values of
\be{muc:'}
\mu_c = \frac{\beta'_\times}{\beta'_\parallel}
\ee
and $\gamma$ satisfies the inequality in~\eqref{bound1}.

Next, we write $p_f$ and $p_{\parallel} \equiv p_\parallel(p_f)$ using the initial values.
From Eq.~\eqref{pfpc2}, $p_f$ becomes
\be{pfpc3}
p_f= p_0 ( 1+e_* G+ \mu_c \tan \phi) , \qquad p_0 \equiv  \frac{m \vi \cos \phi}{(1-\gamma) \beta_\perp'}.
\ee
In addition, from Eq.~\eqref{pf parl}, we have
\be{pf parl2}
p_{\parallel f} 
= \mu_c p_0\left[(1+ e_*G) + \frac{ \mu_c\tan \phi}{\gamma} \right] .
\ee

The velocity of the ball after the collision is, from $ V^0_{\perp} = p_{f}/M,\,  V^0_{\parallel} = p_\parallel (p_f)  /M$,
\be{V0 after}
V^0_{\perp} = \frac{p_0}{M} ( 1+e_* G+ \mu_c \tan \phi), \qquad 
V^0_{\parallel} =  \frac{ \mu_0 p_0}{M} \left[ 1+e_* G+ \frac{\mu_c \tan \phi}{\gamma}\right].
\ee
Integrating Eq.~\eqref{domega dOmega} we get the rotational angular velocity of the ball, 
\be{Omega0 after}
R\vec{\Omega}^0 = \frac{\mu_c p_0}{\nu M} \left[ 1+e_* G+ \frac{\mu_c \tan \phi}{\gamma}\right] \hat z .
\ee

After the impact, let the angle between the velocity of the ball and the $y$-axis be $\theta$. 
Then, from Eq.~\eqref{theta f},
\be{V:angle}
\tan \theta =\frac{V^0_{\parallel}}{V^0_{\perp}}
= \mu_c\Big(1 +
	 \frac{(\gamma^{-1}-1)\mu_c \tan \phi }{1+e_* G+ \mu_c \tan \phi } \Big) .
\ee
The squirt angle, which is the angle of the ball after the impact deviates from the stroke direction of the ball is 
\be{squirt:psi}
 \psi = \phi- \theta_f .
\ee

\vspace{.3cm}
In the previous section, we find that the squirt angle is not proper when we assume the cue is a rigid body. 
Now, let us see how the flexibility of the cue modifies the results. 

To consider the most dramatic situation first, let us observe the limit
$$
\epsilon =\frac{m_c}{M}  \to 0,
$$
i.e., the mass of the cue-end goes to zero. 
In the limit, using $m/m_c \approx 1-\epsilon,$ and $m_c/M_c = \epsilon/Q_0$, one can easily get 
\bea 
\beta_\parallel' &\approx&\beta^0_{\parallel} \equiv \frac{1+\nu_c}{\nu_c}\cos^2 \phi ,
 \nn \\
\beta_\perp' &\approx & \beta^0_\perp \equiv \frac{1+\nu_c}{\nu_c} \sin^2\phi , \nn \\
\beta_\times' &\approx & \beta^0_\times \equiv \frac{1+\nu_c}{\nu_c}\cos \phi\sin\phi    \label{epsilon0 1} .
\eea
Then,  
\be{muc gamma}
\mu_c \to \tan \phi, \qquad \gamma \to  1 
\ee
and the modification factor $G$ of COR becomes
\be{G 0th}
 G \to \frac1{\cos^2 \phi}  .
 \ee 
The bound for the angle~\eqref{bound1} simply gives $\phi < \frac{\pi}4 = 45^{\rm o}$.
This restriction is meaningful because it contains the Coriolis's limit $b/R = 0.7 < \sin 45^{\rm o} \approx 0.707$.
Putting these results to Eq.~\eqref{V:angle} which represents the direction of the ball after the impact, we get $
\tan \theta = \tan \phi.$
Therefore, in this limit the squirt is absent:
\be{psi=0}
\psi = \phi - \theta =0 
\ee
In other words, in the limit of zero-end-mass, the squirt disappears.
To know the velocity of the ball after the impact, let us examine Eq.~\eqref{V0 after}.
Then, we find that the absolute size of  $p_0 = m U_i \cos\phi/(1-\gamma) \beta'_{\perp}$ is ill-defined because both the numerator and the denominator go to zero  in the limit $\epsilon \to 0$.
Explicitly, $m=M \epsilon (1+\epsilon)^{-1} \to 0$ and $\gamma \to 1$.

Therefore, to find the zeroth order velocity of the ball, we need to know the 
results up to the first order in $\epsilon$. 
The mass of the ball is roughly $M = 210 \mbox{ g}$ which gives
$$
\epsilon \equiv \frac{m_c}{M} \approx \frac{12}{210} = \frac{2}{35} \ll 1 .
$$ 
This justifies the perturbative calculation. 
Let us calculate the quantities to their first order to get the velocity.  
First, we write $\beta_k'$ to the first order of $\epsilon$:	
\bea 
\beta_\parallel' &\approx& \beta^0_\parallel (1-\epsilon A_\parallel);
\quad
A_\parallel \equiv 
	1- \frac{\nu_c(1+\nu)}{\nu(1+\nu_c) \cos^2\phi} - \frac{\nu_c \tan ^2\phi}{Q_0(1+ \nu_c)}
	 , \nn \\
\beta_\perp' &\approx& \beta^0_\perp  (1-\epsilon A_\perp); \quad
A_\perp \equiv 1- \frac{\nu_c}{(1+\nu_c) \sin^2\phi} 
	-\frac{\nu_c \cot ^2\phi}{Q_0(1+ \nu_c)}
, \nn \\
\beta_\times' &\approx & \beta^0_\times (1-\epsilon A_\times);\quad
A_\times\equiv  1+ \frac{\nu_c}{(1+\nu_c) Q_0} .
\eea
Here, $Q_0= M_c/M$ represents the mass ratio between the cue and the ball.
Putting these into Eqs.~\eqref{muc:'} and \eqref{gamma:'} we get
$$
\mu_c \approx \tan \phi \,[1-\epsilon(A_\times-A_\parallel)] = \tan \phi \left[1- \frac{\epsilon \nu_c}{(1+\nu_c) \cos^2\phi} \left(\frac1{Q_0} + \frac{1+\nu}{\nu} \right) \right],
$$
and
$$
\gamma \approx 1- \frac{\epsilon \nu_c}{1+\nu_c} \frac1{\sin^2\phi \cos^2\phi}
\left[1+\frac1{Q_0} +\frac{\sin^2\phi}{\nu}  \right] .
$$
Putting these results to Eq.~\eqref{V:angle} to obtain the angle between the velocity of the ball and the $y$-axis, we find
\bea \label{theta f: 1}
\tan \theta \approx  \tan \phi 
\left\{1- \frac{\epsilon \nu_c}{(1+\nu_c)(1+ e_*)} 
	\left[\frac{1}{\nu}+ \left(\frac{1}{\nu}+1+\frac1{Q_0}\right) 
	   \frac{e_*}{\cos^2\phi}
	\right]
\right\} .
\eea
Therefore, the squirt angle $\psi$ to the first order in  $\epsilon$, by using the formula $\psi \approx \tan \psi = \tan (\phi - \theta)$, becomes
\bea \label{psi squirt}
\tan \psi \approx
 \frac{\epsilon \nu_c  \sin\phi \cos \phi }{(1+\nu_c)(1+e_*)} \left[ \frac1{\nu}+
 	\left(\frac{1}{\nu}+1+\frac1{Q_0}\right) \frac{e_*}{\cos^2\phi} 
\right] .
\eea
As you can see from this equation, the squirt angle depends on various factors.
For example, $\nu_c$ is a value related to the structure of the cue-end, $\nu$ is dependent on the mass distribution of the ball, and $e_*$ is the COR for the frontal collision between the cue and the ball.
Also, $Q_0$ corresponds to the mass ratio between the cue and the ball.
The squirt angle decreases as this mass ratio increases.
That is, the squirt angle decreases as the mass of the cue increases with fixed end-mass.
However, as can be seen by comparing with the number $(1+\nu)/\nu \sim 7/2$, if the mass ratio is higher than a certain value (approximately $Q_0 \geq 3$), the effect of this mass ratio is not significant. 
On the other hand, if the mass ratio decreases, this squirt angle can be noticeably increased, especially if the cue mass is less than that of the ball.
Finally, the squirt angle simply proportional to $\epsilon = m_c/M$, the mass ratio between the cue-tip and the ball.
 \begin{figure}[htb]
\begin{center}
\begin{tabular}{cc}
 \includegraphics[width=.50\linewidth,origin=tl]{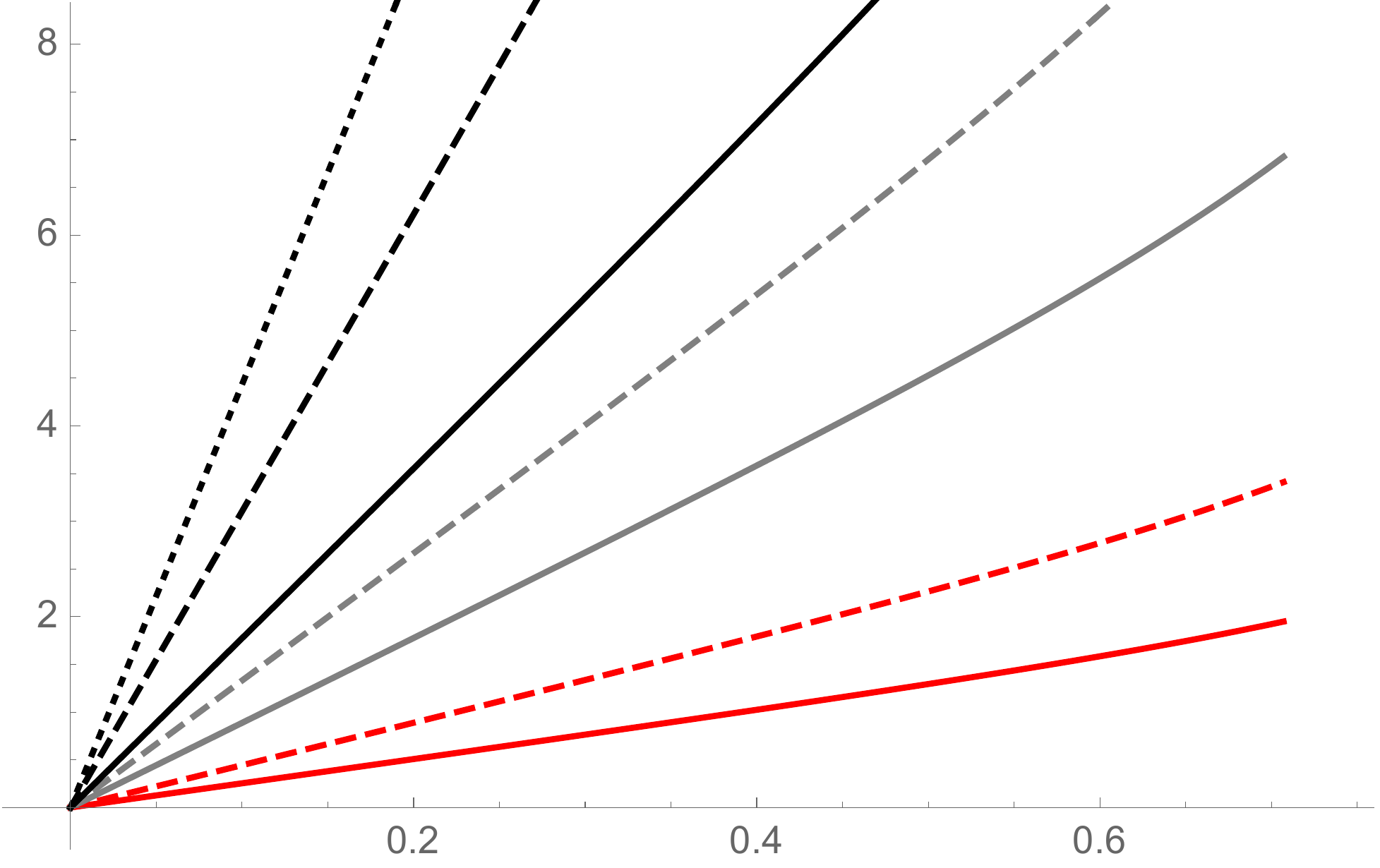} &
  \includegraphics[width=.50\linewidth,origin=tl]{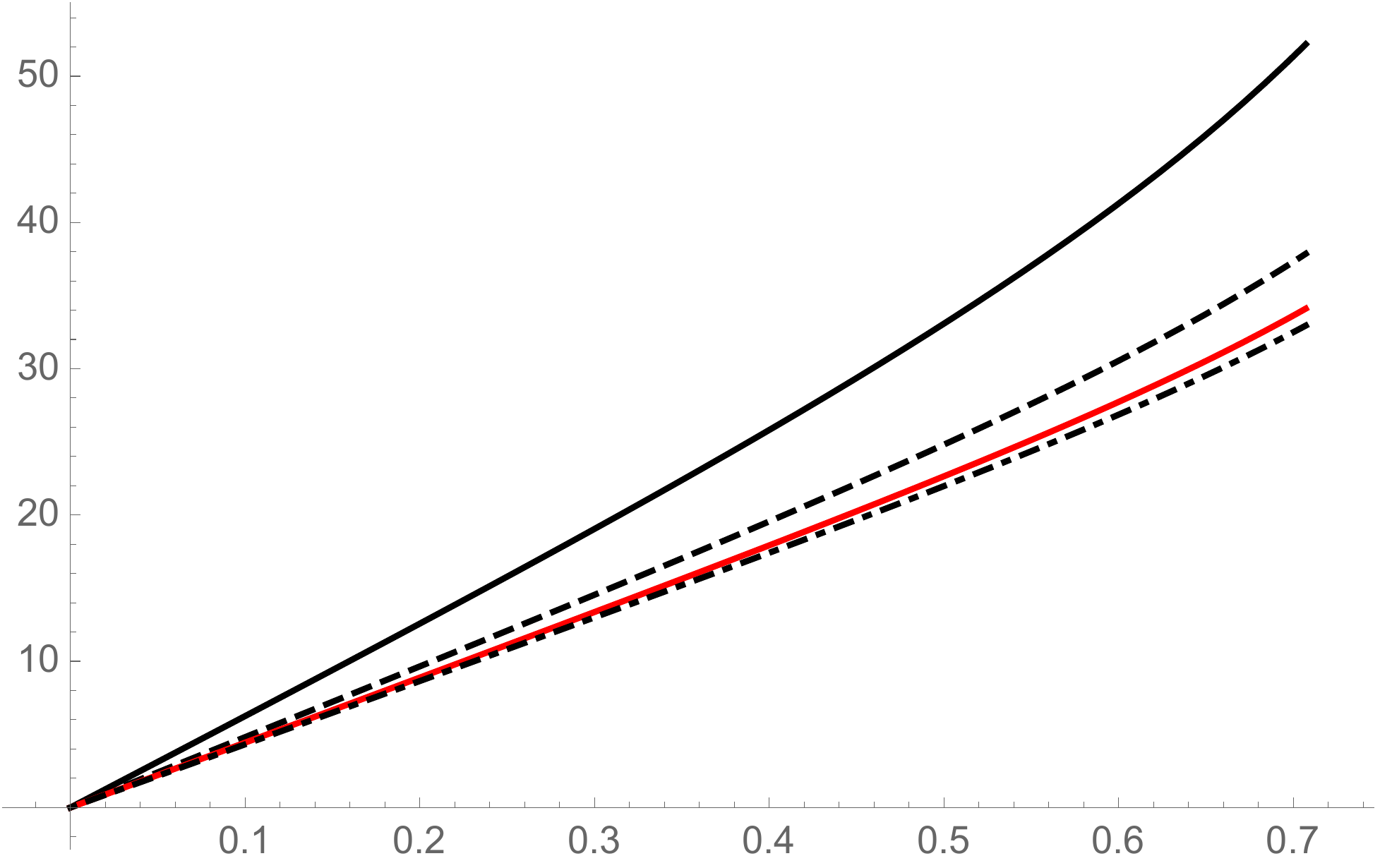} 
\end{tabular}
\put (-518,72) {$ \psi $ }
\put (-263,-75) {$ \frac{b}{R} $  }%
\put (-10,-75) {$ \frac{b}{R} $  }
\put (-253,78) {$\displaystyle \frac{\psi}{\epsilon} $ }
\end{center}
\caption{The squirt angle for a realistic model of cue. 
For the left panel, $m_c/M = 0, 2/35, 0.1, 0.2, 0.3, 0.4, 0.7, 1.0$ respectively from the bottom. 
The solid red line corresponds to $m_c/M = 2/35$. 
Here, we choose the typical values for COR, $e_*=3/4$, and cue-ball mass ratio, $Q_0 = 18/7$. 
For the right panel we plot $\psi/\epsilon$ with respect to $b/R$ for, $M_c/M =0.3,~1,~18/7, 5$, respectively from the top.  }
\label{fig:squirt}
\end{figure} 
In the left panel of Fig.~\ref{fig:squirt}, we plot the squirt angle for several values of $\epsilon$ for $Q_0 = 18/7$ and $e_* = 3/4$.
It is obvious that the squirt angle is proportional to $\epsilon$, the mass ratio between the ball and the cue-end. 
In addition, one can notice that the squirt angle increases almost linearly for small impact parameter $b$. 
For larger impact parameter $b \sim .5$, non-linear behaviors just appears. 

For a well-made cue, $(m_c/M <0.1)$, if the impact parameter is less than $0.55 R$, the squirt angle is less than 1 degree.
With a squirt angle of 1 degree, in the case of the carom billiards table, an error of approximately 49.6 mm occurs when we strike the ball from one end to the other end in the long axis direction.
Since the radius of the billiard ball is about 36.9 mm, this makes an error of about 5/8 thickness.
In playing billiards, personal cues are especially meaningful because the player notices this squirt angle of his own cue.

In 2001, Ron Shepard~\cite{Shepard} calculated the squirt angle by using the law of momentum conservation after assuming that the terminal velocity of the cue-end is the same as the terminal velocity of the impact point\footnote{If one can ignore the rotational motion of the cue tip, the translational velocity of the cue-end and the final speed of the impact point will be the same.}.
If this assumption is correct, we can write the squirt angle in an easier form,
$$
\tan \psi \approx \frac{ \frac{b}{\nu R} \sqrt{1- (b/R)^2}}{1+ 1/\epsilon + (b/R)^2/\nu }.
$$
This formula presents a result similar to that in Fig.~\ref{fig:squirt} obtained in this work qualitatively regarding the mass-ratio between  the cue-end and the ball.
However, it gives very different numerical values. 
For example, if $\epsilon = 0.1$, this formula gives the squirt angle of 4.38$^{\rm o}$, but the formula in Eq.~\eqref{psi squirt} gives a squirt angle smaller than $1^{\rm o}$.

The present work allows to take into account the effect of the cue-ball mass ratio, the effect of the change of the elasticity, and the effect of the structural change of the cue-tip on the squirt angle.

\section{Summary: The state of the ball and the cue after the impact}

In this work, we have studied the impact of a cue on a billiard ball. 
Respecting the flexibility of a cue, the cue is modeled to be composed of two parts: the body and the cue-end. 
The cue-end is the short `end-part of the cue' which responses to the impact from the ball instantly. 
The mass of the cue-end is much lighter than the mass of the ball.
We find that the squirt angle is suppressed to $O(\epsilon)$ where $\epsilon = m_c/M$, the mass ratio between the cue-end and the ball. 
The squirt angle increases with the impact parameter $b = R \sin \phi$, where $R$ denotes the radius of the ball. 
For small $b \ll R$, it increases linearly.
For $b\sim R/2$, non-linear behavior start to appear, which can be seen in the experimental data in~\cite{Cross2008}.

Before finishing this work, we summarize the motion of the ball struck by the cue with the impact parameter $b= R\sin\phi$. 
This is very important for the stroke of the billiard ball but I cannot find the results in the literature.
We strike the cue horizontally on the height of the center of the ball. 
Because the squirt angle is very small we may use the approximation $\epsilon \ll 1$ and write the velocity only to the first non-vanishing order, from Eq.~\eqref{pfpc3},
$$
\frac{p_0}{M} \approx \frac{ U_i \cos^3 \phi}{(1+ Q_0^{-1}) +\nu^{-1}\sin ^2 \phi} .
$$
Then, the translational velocity of the ball, by using Eqs.~\eqref{muc gamma}, \eqref{G 0th}, is
\be{Vf: 0}
\vec{V}^0 = \frac{1+e_*  }
{1+ Q_0^{-1} +\nu^{-1}\sin^2 \phi} \vec{U}_i  .
\ee
Here, $\vec{ U}_i$ denotes the speed of the cue before the impact. 
Therefore, most importantly, the velocity of the ball increases with the initial velocity of the cue. 
The velocity is maximized at $\phi =0$.
The rotational velocity of the ball is 
\be{ROmega: 0}
R\vec{\Omega}^0 = \frac{U_i \sin\phi}{\nu} \frac{1+ e_* }{1+ Q_0^{-1} + \nu^{-1}\sin ^2 \phi}   \hat z .
\ee
The magnitude of this angular velocity satisfies $R |\Omega| = V_\parallel/\nu$.
Note that the size of this angular velocity takes its maximum value 
\be{Omega max}
\left|R\Omega^0\right|_{\rm max} =  \frac{U_i }{2\sqrt{\nu}} \frac{1+ e_* }{\sqrt{1+ Q_0^{-1}}} 
\ee
at
$$
\sin \phi_M = \sqrt{\nu (1+ Q_0^{-1})} .
$$
As the mass ratio $Q_0$ increases, the angle $\phi_M$ where the angular velocity takes its maximum decreases. 
However, for an ordinary cue-ball mass ratio such as $Q_0 = 18/6$ or $Q_0=3$, the angle becomes $\sin\phi_M = 0.745$ or $\sin \phi_M = 0.730$, respectively. 
Both are located outside of the bound of the impact parameter $b= R \sin\phi_M > b_{\rm max} = 0.7 R$.
Therefore, within the bound of $\phi$, the angular velocity increases with $\phi$. 
If the cue is massive so that $Q_0^{-1} \ll 1$, we have $\sin\phi_M = \sqrt{2/5} \sim 0.632$.
In this case, the impact parameter corresponding to $\phi_M$ is located within the range of the bound $b=0.7R$. 
Therefore, one may find that the angular velocity achieved by the stroke decreases with $\phi > \phi_M$. 

This relation holds generally for every stroke performed horizontally if an artificial change of force from the hand is absent.
The angular velocity~\eqref{ROmega: 0} is obtained by striking the cue on a position with the same height as the ball's center.
If the cue mass is minute, one can find that both the translational velocity and the spin of the ball are small.
On the other hand, for a higher cue mass, a faster translational speed and spin are achieved, even though they cannot increase indefinitely.
An interesting observation is that the rotational angular velocity thus obtained never depends on the magnitude of the friction coefficient between the cue and the ball.
In other words, the friction coefficient determines only the limiting value that prevents the cue-miss.

Let us find the rotational angular velocity when a cue strikes the location of the impact point $\vec{R}_I = (h, y,\ell-R)$ with the impact parameter $b$.
The simplest way to do this job is i) to rewrite the above rotational angular velocity by using the impact parameter, the vector from the center of the ball toward the impact point, and the vector representing the direction of the cue stroke, and then ii) generalize the equation.
This is possible because the period the cue strikes the ball is so short that neither gravity nor the friction between the ball and the table plays a role.
First, the impact parameter satisfies
$$
b = \sqrt{h^2 + (\ell -R)^2}, \qquad  y = \sqrt{R^2 -b^2}
$$ 
At the present case of the previous section, $\ell = R$, $h=b$.

Let the unit vector along the impact direction of the cue be $\hat{V}_{\rm cue}$.
Then, the rotational direction of the ball can be written as
$$
\hat z = \frac{\vec{R}_I \times \hat{V}_{\rm cue}}{b} .
$$

The total kinetic energy of the ball, summing the kinetic translational energy and the rotational energy, is
$$
K=  \frac{MU_i^2 }{2} \left(\frac{1+ e_* }{1+ Q_0^{-1} + \nu^{-1}\sin ^2 \phi}  \right)^2
 	\left(1+ \frac{\sin^2\phi}{\nu} \right) .
$$
The kinetic energy of the ball monotonically decreases with $\phi$. 
Therefore, the efficiency of kinetic energy transferal from the cue to the ball happens only when the head-on collision.

The velocity of the cue, by integrating Eq.~\eqref{dW}, is $\vec{V}_{\rm cue} = \vec{\vi}+\int d \vec{V}_{\rm cue}$.
 The direction of $\vec{V}_{\rm cue}$ is always parallel to $\vec{\vi}$.
 Then, the speed of the cue after the collision becomes
\bea
V^f_{\rm cue} &=& \vi- \frac{\cos \phi \, p_\perp + \sin \phi \,p_\parallel}{M_c} \nn \\
&=&\vi-\frac{p_0}{M_c} \left[ ( 1+e_* G+ \mu_c \tan \phi) + \mu_c\left( 1+e_* G+ \frac{\mu_c \tan \phi}{\gamma}\right)  \right] \nn \\
&=&\vi\left[1-\frac{ ( \cos \phi +\sin \phi) \left( 1+e_*\right)}{(1+ Q_0^{-1}) +\nu^{-1}\sin ^2 \phi} \right]  .\label{Vf: cue}
\eea

The angular velocity of the cue-end can be obtained by integrating Eq.~\eqref{domega dOmega}. 
Using Eq.~\eqref{pfpc3} and \eqref{pf parl2}, we get
\bea 
\omega_z &=& \frac{r_c}{I_c} (\cos \phi \, p_\parallel - \sin \phi \, p_\perp)  \nn \\
&=& \frac{U_i \cos^2 \phi}{(1+\epsilon)\nu_c r_c (1-\gamma) \beta_\perp'} \left[ (1+ e_* G) (\mu_c -\tan \phi)
	+\mu_c \tan \phi \Big(\frac{\mu_c}{\gamma} - \tan \phi\Big) \right]   \nn\\
&\approx& - \frac{U_i \cos \phi \sin\phi}{(1+\nu_c)r_c} 
\frac{(1+ \frac{e_*}{\cos^2\phi} ) \left(\frac1{Q_0}+1+\frac1{\nu}\right)
	-\tan \phi \Big( 1+\frac1{Q_0} \Big) }
{1+\frac1{Q_0}+\frac{\sin^2\phi}{\nu} }  .
\eea	
The rotational velocity increases with $\phi$ even though its direction is along the negative direction of $\hat z$. 
What is important here is that the angular velocity of the cue-end is finite.
Therefore, the rotational energy of the cue-end, in the  $\epsilon \to 0$ limit, goes to zero:
$$
E_{\rm cue-end, rotation} = I_c \omega_z^2  \to 0 .
$$
It should be noted that this rotational angular velocity represents the rotation of the cue-end rather than the rotation of the cue itself.
 
In this calculation, the squirt was considered to be very small.
Of course, when striking the ball to send it a distant target, even a small squirt angle of $0.5^{\rm o}$ creates an error of about half a ball, which we cannot ignore.
The squirt angle of the ball is very small, so we do not write it down. 
Please refer the formula ~\eqref{psi squirt} or Fig.~\ref{fig:squirt}.
 
Consider the case that a player holds a cue tightly when he strikes a billiard ball. 
He does not apply extra-force at the instance of the cue striking the ball. 
Because the arm moves with the cue, this is equivalent to an effective increase of the cue's mass, which makes the parameter $Q_0$ increase. 
Then, from Eqs.~\eqref{psi squirt}, \eqref{Vf: 0}, and \eqref{ROmega: 0}, the velocity and the angular velocity of the ball after the impact increase but the squirt angle tends to decrease even though this effect is not significant. 
As $Q_0$ increases, the maximum angular velocity in Eq.~\eqref{Omega max} also increases.  
In addition, the maximum value of the angular velocity can be achieved for smaller impact parameter. 

\section*{Acknowledgment}
This work was supported by the National Research Foundation of Korea grants funded by the Korea government NRF-2020R1A2C1009313.
The author thanks to Youngone Lee for helpful discussions.
 

\end{document}